\documentclass[12pt, preprint]{aastex}
\begin{document}
\title{AN OBSERVATIONAL SIGNAL OF THE VOID SHAPE CORRELATION AND ITS LINK TO THE COSMIC WEB}
\author{Jounghun Lee}
\affil{Astronomy Program, Department of Physics and Astronomy, 
Seoul National University, Seoul 151-742, Korea}
\email{jounghun@astro.snu.ac.kr}
\author{Fiona Hoyle}
\affil{Pontifica Universidad Catolica de Ecuador, 12 de Octubre 1076 y Roca, Quito, Ecuador}
\email{fionahoyle11@gmail.com}
\begin{abstract}
The shapes of cosmic voids are prone to distortions by the external tidal forces since their low-densities imply a lower internal 
resistance. This susceptibility of the void shapes to the tidal distortions makes them useful 
as an indicator of the large-scale tidal and density fields, despite the practical difficulty in defining them. 
Using the void catalog constructed by \citet{pan-etal12} from the Seventh Data Release of the Sloan 
Digital Sky Survey (SDSS DR7), we detect a clear $4\sigma$ signal of spatial correlations of the void 
shapes on the scale of $20\,h^{-1}$Mpc and show that the signal is robust against the projection of the void shapes onto the 
plane of sky. By constructing a simple analytic model for the void shape correlation,  within the framework of tidal torque theory, 
we demonstrate that the void shape correlation function scales linearly with the two-point correlation function of the linear 
density field. We also find a direct observational evidence for the cross-correlation of the void shapes with the large-scale 
velocity shear field that was linearly reconstructed by Lee et al. (2014) from the SDSS DR7.  We discuss the possibility of 
using the void shape correlation function to break the degeneracy between the density parameter and the power spectrum 
amplitude and to independently constrain the neutrino mass as well.  
\end{abstract}
\section{INTRODUCTION}
\label{sec:intro}

The standard statistics of the large scale structure (LSS) such as the matter density power spectrum, abundance evolution of the 
galaxy clusters, galaxy bias function, and redshift distortion effect have been prevalently used to probe the initial conditions of the 
universe. Although the power of the standard LSS statistics as a cosmological probe has been confirmed by numerous 
observational and numerical studies, its limitation has recently been recently pointed out. For instance, 
\citet{wei-etal13} have recently proved that the standard LSS statistics inherently fails to distinguish between two different cosmological 
models: coupled dark energy (cDE) and modified gravity (MG)  models. In the former the cosmic acceleration is caused by the 
anti-gravitational effect of a scalar field dark energy coupled to dark matter \citep[see][for a review]{AT10} 
while the latter attributes the apparent acceleration of space-time to the deviation of gravity from the General Relativity on cosmological 
scale \citep[see][for a review]{MG_review12}. 

To overcome the limitation of the standard LSS statistics, it is necessary to develop as many nonstandard complimentary probes 
as possible. The distribution and correlation of void shapes is one of those nonstandard LSS statistics that has arisen  
in recent years as a complimentary probe of cosmology \citep[e.g.,][]{PL07a,platen-etal08,LP09,LW10,biswas-etal10,bos-etal12}. 
The classical work of \citet{icke84} predicted that the shape of an isolated void region would become rounder and rounder as its 
expansion grows faster than the expansion of the rest of the universe. 
However, it was later pointed out that the cosmic voids should be regarded not as isolated low-density regions but as being in constant 
interaction with the surrounding matter distribution \citep{VK93,dubinski-etal93,sahni-etal94}. Given that the internal resistance of a 
region to the tidal distortion diminishes with its density, \citet{shandarin-etal06} suggested that the void shapes should gradually develop 
nonspherical asymmetry under the tidal shear influence from the surrounding matter distribution. 

The statistical link between the non-spherical shapes of cosmic voids and the surrounding tidal shear field have been analytically and 
numerically investigated by several authors.  Noting that the void shapes become most (least) stretched along the directions of the 
minimum (maximum) external tidal force and that the spatial correlations of the tidal fields would generate the spatial corrections of the 
the void shapes, \citet{LP06} introduced a concept of a void spin angular momentum vector to quantify the direction of 
an asymmetric void shape. Measuring the spatial correlations of the spin axes of the voids identified in a $N$-body simulation, they 
found that the numerical result of the void spin correlation matched the analytic prediction based on the tidal torque theory. 
\citet{PL07b} found strong alignments between the void spin axes and the elongated axes of the neighbour superclusters identified 
in a $N$-body simulation and interpreted this as a numerical evidence for the tidal influences on the voids shapes, claiming that 
that the shapes of the superclusters would be elongated along the directions of the minimum tidal forces.

Although the new concept of void spin effectively quantifies the influence of the anisotropic tidal field on the void shape, as shown in the 
pioneering work of \citet{LP06}, the void spin axis is hard to measure in practice since it requires knowledge of the 
peculiar velocities of the void galaxies. 
By quantifying the orientations of the asymmetric void shapes by the distinct principal axes of their inertia momentum, rather than their 
spin axes,  \citet{platen-etal08} measured the spatial correlations between the axes of the asymmetric void shapes, by analyzing the data 
from a $N$-body simulation. They showed that the spatial correlations between the void shapes decrease slowly with the separation 
distance, exhibiting significant correlation strength even on scales larger than $30\,h^{-1}$Mpc.   

Yet no direct observational evidence of the tidally induced correlations of the void shapes has been reported.  The main difficulty  
in dealing with the void shapes is its over-sensitivity to the void-identification scheme.  Although a variety of the void-finding 
algorithms have been compared with one another in the comprehensive work of \citet{colberg-etal08}, it is still a touchy issue 
to decide which void finding algorithm is optimal since each algorithm has its own merits and limitations in 
defining the voids and characterizing their properties. Furthermore, the relatively low abundance of observed voids in the local 
universe hinders performing statistics of void shapes where large numbers of voids are required \citep[e.g.,see][]{bos-etal12}. 
Notwithstanding these difficulties, an observational confirmation of the theoretical prediction for the link between the void shapes and the 
tidal shear field is essential to see if statistics of void shapes can be used as a complimentary test of cosmology. 
The goal of this Paper is to find a direct observational evidence for the tidally induced correlations of the non-spherical void shapes 
from recently available large galaxy surveys. 

The organization of this Paper is as follows. In section \ref{sec:signal}, we report a high signal detection of void shape 
correlations. In section \ref{sec:model} the observational signals are compared with a simple analytic formula
based on the tidal torque theory. In section \ref{sec:shear} we present the first direct observational evidence for the cross-correlations 
between the void shapes and the large-scale tidal field. In section \ref{sec:summary} the implications and future applications 
of our final results are discussed. Throughout this Paper, the key cosmological parameters assume the values of  
$\Omega_{\Lambda}=0.26,\ \Omega_{\Lambda}=0.74,\ h=0.73,\ \sigma_{8}=0.8$, for a universe dominated by the cosmological 
constant ($\Lambda$) and cold dark matter (CDM) with Euclidean geometry.

\section{DETECTION OF  A SIGNAL OF VOID SHAPE CORRELATIONS}\label{sec:signal}

To find an observational signal of the void shape correlation, we analyze the void catalog constructed by \citet{pan-etal12} who identified  
the low-$z$ voids by applying the void-finding algorithm of \citet[][HV02 hereafter]{HV02} to the Seventh Data Release of 
the Sloan Digital Sky Survey (SDSS DR7) \citep{sdssdr7}.   
Selecting only those voids with $30$ or more member galaxies from the void catalog of \citet{pan-etal12}, 
we construct a sample of $831$ giant voids whose effective radius  ($R_{\rm v}$) and number of the member galaxies ($N_{\rm vg}$) 
are in the ranges of $9.92\le R_{\rm v}\le 33.92$ and $30\le N_{\rm vg}\le 2984$, respectively.

It may be worth discussing here the reason for using the galaxy voids identified by the HV02 algorithm rather than those by the more 
recent and elaborate void finders like ZOBOV \citep{zobov}. First of all, the HV02 algorithm can be readily implemented into the spatial 
distributions of the observed galaxies from the large-scale survey to identify the voids, without making any assumption 
about the galaxy bias or the halo abundances \citep[see also,][]{HV04, pan-etal12} . 
Second of all, we intend to compare the observed signal of the void shape correlations with an analytic model, constructed as 
a modification of the void spin correlation model of \citet{LP06} (see section \ref{sec:model}). When \citet{LP06} derived their 
analytic model for the void spin correlation function, they tested it against the numerical result that had been obtained by utilizing 
the HV02 algorithm to a $N$-body simulation and found a good agreement, which motivated us to select the HV02 algorithm 
for the current analysis. 
 
Using information on the redshifts, right ascensions and declinations, we first measure the positions of the member galaxies 
of each void and then determinate the location of its luminosity center, ${\bf x}^{0}=(x^{0}_{i})$, as
\begin{equation}
{\bf x}_{\rm 0} = \frac{\sum_{\alpha} M_{\rm r}^{\alpha}{\bf x}^{\alpha}}{\sum_{\alpha} M_{r}^{\alpha}}\ . \ 
\end{equation}
where ${\bf x}^{\alpha}$ is the position of the $\alpha$-th galaxy in a given void and $M_{r}^{\alpha}$ is its absolute magnitude in 
the $r$-band. The inertia momentum tensor of each void, ${\bf I}=(I_{ij})$, is now calculated as \citep{LP06}:
\begin{equation}
\label{eqn:void_iner}
I_{ij} = \sum_{\alpha}M^{\alpha}_{r}\left(x^{\alpha}_{i}-x^{0}_{i}\right)\left(x^{\alpha}_{j}-x^{0}_{j}\right)\, ,
\end{equation}
Diagonalizing ${\bf I}$ for each void, we find its three (orthonormal) eigenvectors (say, ${\bf u}_{1},\ {\bf u}_{2},\ {\bf u}_{3}$) 
corresponding to the three eigenvalues ($I_{1},\ I_{2},\ I_{3}$ in a decreasing order) and label them as the major, 
intermediate, and minor principal axes of the void shape, respectively. 

Note that the {\it void shape} in the current work is defined in terms of the spatial positions of the galaxies in a void region relative 
to their luminosity center, which is different from the true geometry of a void region. Since the void galaxies are located 
in relatively denser section of the void region, their luminosity center would deviate from the density minimum of the void region 
that is usually defined as the void center. 
Although the spatial distributions of the void galaxies should develop anisotropy under the influence of the external tidal shear field, the 
non-spherical geometry of a void region with density minimum as its center should be a better indicator of the external tidal field 
than the anisotropic spatial distribution of the void galaxies. Unfortunately, however,  it is very hard to accurately 
measure the three dimensional geometry of a void region from observational data since its boundary depends very sensitively 
on which void finding algorithm is used \citep{colberg-etal08}. To avoid this practical difficulty and to be consistent with 
\citet{LP06}, we use the anisotropic spatial locations of the void galaxies to define the void shape and call their luminosity center 
the void center throughout this paper.

\begin{figure}[tb]
\includegraphics[scale=1.0]{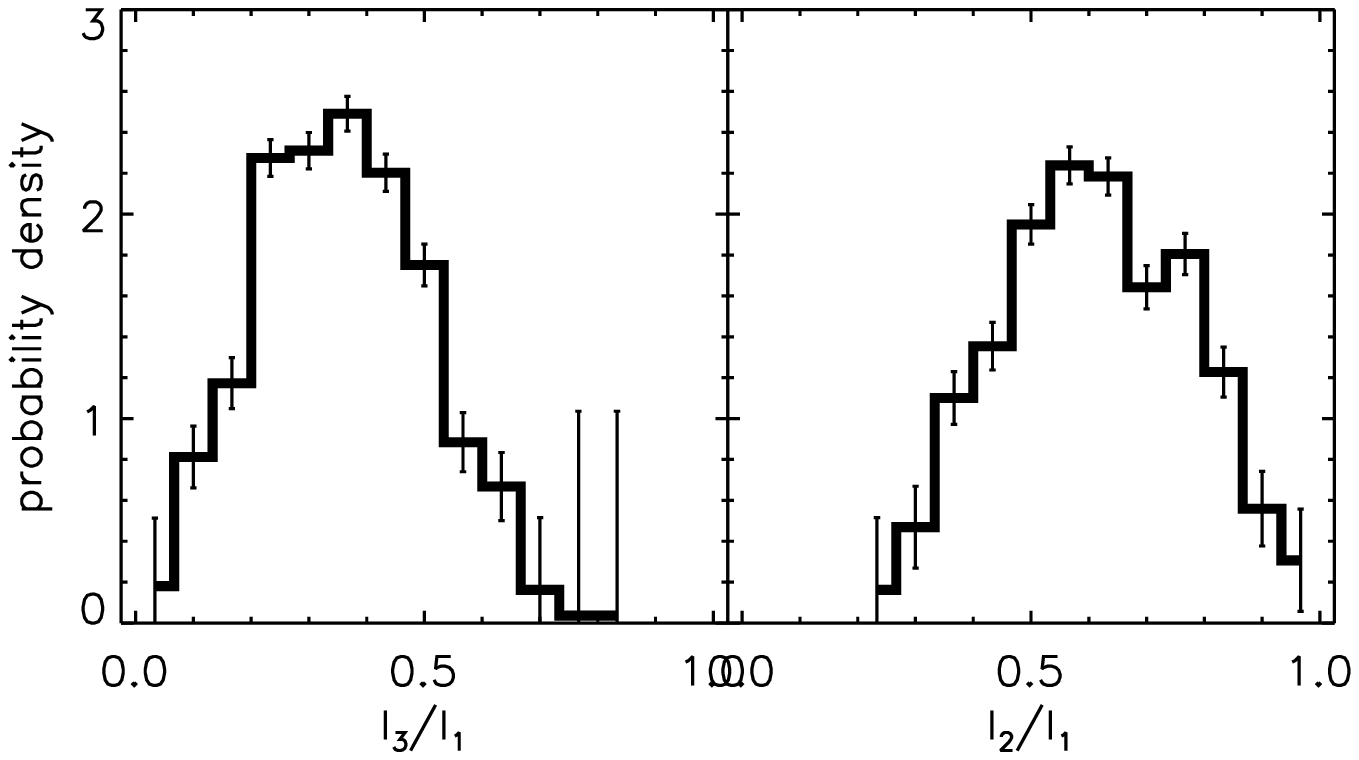}
\caption{Probability density distributions of the ratios of the smallest (second smallest) eigenvalues to the largest ones of the 
inertia tensors of the sample voids from the SDSS DR7 with Poisson errors in the left (right) panel. }
\label{fig:ratio}
\end{figure}

If a void had a spherically symmetric shape, then three eigenvalues of its inertia momentum tensor would be identical and thus 
the corresponding three eigenvectors would be degenerate. The more nonspherical the shape of a void is, the more distinct three 
eigenvectors of its inertia momentum tensor are from one another. 
Measuring two ratios of $I_{2}/I_{1}$ and $I_{3}/I_{1}$ for each void, we find the probability density distributions, $p(I_{2}/I_{1})$ 
and $p(I_{3}/I_{1})$, which are plotted as histograms with Poisson errors in the left and right panels of Figure \ref{fig:ratio}, 
respectively.  As can be seen, the two distributions, $p(I_{3}/I_{1})$ and $p(I_{2}/I_{1})$ attain their maxima at 
$I_{3}/I_{1}\approx 0.4$ and $I_{2}/I_{1}\approx 0.6$, respectively, which shows that three principal axes of the voids in the 
sample are quite distinct from one another.

The spatial correlation function of the void shapes, $\eta_{\rm v}(r)$, is defined in a similar way as the spatial correlation 
function of the halo shapes as defined by \citet{lee-etal08}:   
\begin{equation}
\label{eqn:void_shape}
\eta_{\rm v}(r) \equiv \langle\vert{\bf u}({\bf x})\cdot{\bf u}({\bf x}+{\bf r})\vert^{2}\rangle - \frac{1}{3}\, ,
\end{equation} 
where ${\bf u}({\bf x})$ and ${\bf u}({\bf x}+{\bf r})$ are the principal axes of two voids located at the positions of 
${\bf x}$ and ${\bf x}+{\bf r}$, respectively. In Equation (\ref{eqn:void_shape}) the ensemble average of the first term in the 
left hand side is taken over all void pairs in the sample whose separation distance is $r$, while the second term, $1/3$, represents the 
expectation value of the first term under the null hypothesis of no correlation. Note that $\eta_{\rm v}(r)$ would be zero but for correlation.

To determine $\eta_{\rm v}(r)$, we take the following steps.  First,  we find the range of the separation distances, $r$, 
of all pairs from the $831$ voids in our sample. Second, we divide the values of $r$ into small bins with size $\Delta r$ and 
find the void pairs whose separation distances are in the interval of $[r,\ r+ \Delta r]$.  Third, we compute the dot-products between the 
principal axes of the inertia momentum tensors of two voids in each pair and then take the ensemble average over those void pairs 
whose separation distances belong to each $r$-bin. Forth, we subtract $1/3$ from the ensemble average to obtain $\eta_{\rm v}$ and 
calculate the standard deviation of $\sigma_{\eta}$ in the measurement of $\eta_{\rm v}$ at each $r$-bin. 
Fifth, we estimate the scatters of $\eta_{\rm v}$ among $1000$ bootstrap resamples. 

\begin{figure}
\includegraphics[scale=1.0]{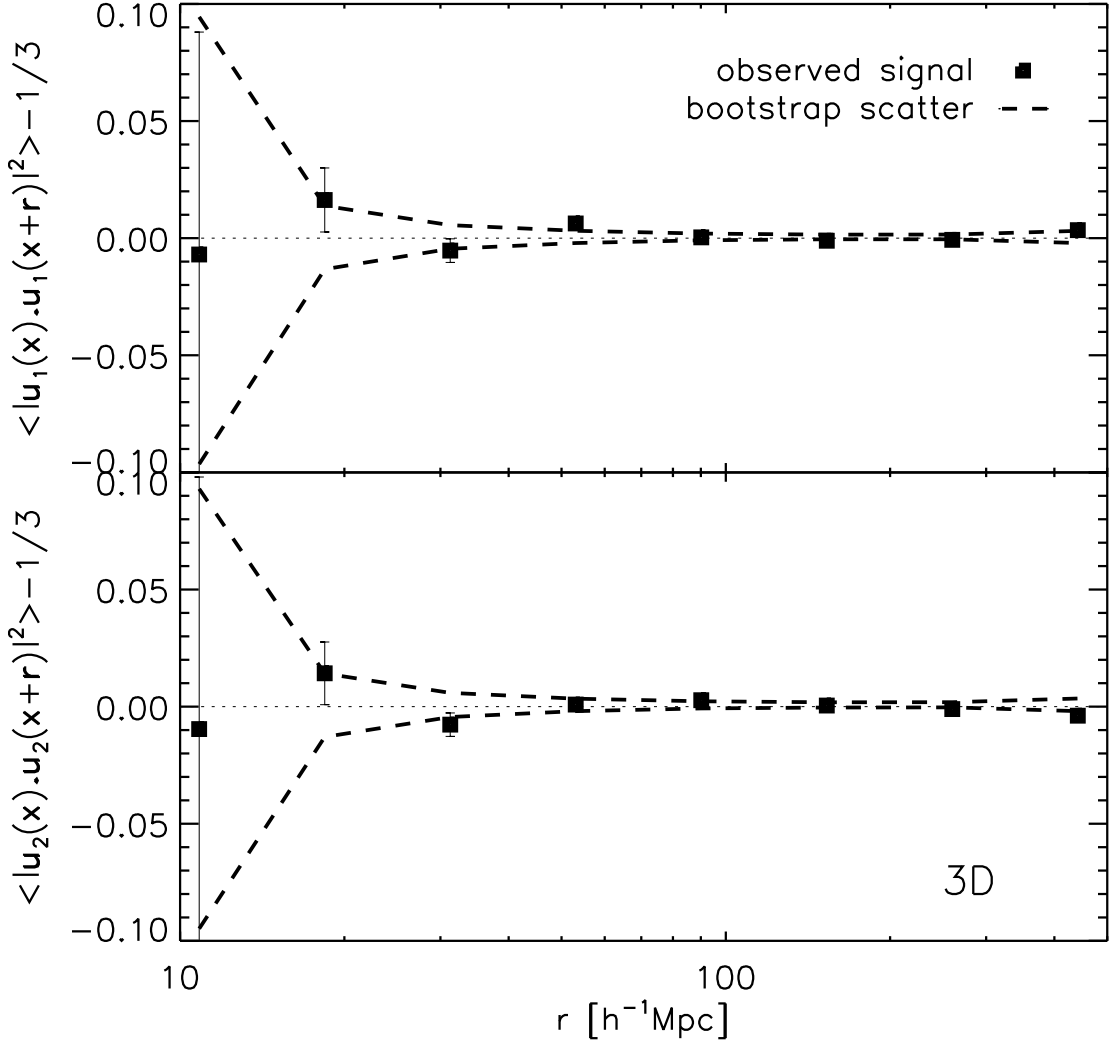}
\caption{Spatial correlations of the major and intermediate principal axes of the inertia tensors of the sample voids as square 
dots in the top and bottom panels, respectively. In each panel, the errors represent the one standard deviation in the measurement 
of the mean correlations, the dotted line corresponds to the case of zero correlation, and the dashed lines show the scatters 
among the bootstrap $1000$ resamples.}
\label{fig:major}
\end{figure}

Figure \ref{fig:major} shows the spatial correlation function of the major and intermediate principal axes of the void inertia 
tensors as square dots with one standard deviation errors $\sigma_{\eta}$ in the top and bottom panels, respectively. In each 
panel, the dotted line represents the case of no spatial correlations and the dashed line depicts the bootstrap scatter among the 
$1000$ resamples. 
As can be seen,  the major and intermediate principal axes of the void inertia tensors show no signal of spatial correlations.
Figure \ref{fig:minor} shows the same as Figure \ref{fig:major} but for the case of the minor principal axes of the void inertia tensors. 
As can be seen, there is a clear signal of spatial correlation as high as $4\sigma_{\eta}$ at the bin of $r\approx 20\,h^{-1}$Mpc. 
At the larger separation distances, the correlation signal is not statistically significant, lower than twice the bootstrap error.

To confirm that the observed signal of the spatial correlations of the minor principal axes of the void inertia tensors is not 
a spurious one generated by some systematic errors involved in the measurements of the galaxy positions in the redshift space, 
we redo the alignment analyses in the two dimensional plane of sky. For each void, we project the minor principal axes of 
its inertia tensor along the line of sight direction to the void center and then determine the spatial correlations of the projected 
minor principal axes as a function of the separation distance, the result of which is displayed in Figure \ref{fig:2d}. Here, 
${\bf u}_{3}^{2d}$ denotes the projected minor principal axes of the void inertia tensors onto the plane of sky. Note that for the 
case of no correlation, the ensemble average terms is expected to yield one half rather than one third for the case of the projected 
axes. As can be seen, a correlation signal is still present with almost the same strength at $r\approx 20\,h^{-1}$Mpc even using the 
projected directions, which verifies that the observed signal of the void shape correlation is truly intrinsic. 

It is worth explaining here why only the minor principal axes of the void inertia tensors show correlations.  Since a void region 
would become most stretched along the direction of the minimum external tidal force, it might seem natural for us to expect that 
the major principal axes of the void inertia tensors should be as strongly correlated as their minor counterparts. 
However, as mentioned in the above, we define the void shape in terms of the spatial positions of the void galaxies relative to their 
luminosity center.  The most stretched section of a void region must be the emptiest, being almost devoid of the galaxies. Thus, when 
the void inertia tensor is defined from the locations of the void galaxies, the major principal axis of the inertia tensor should be a poor 
approximation to the direction of the minimum external tidal force, while its minor axis traces well the direction of the maximum  
external tidal force. In other words, the major principal axes of the void inertia tensors appear to be uncorrelated because they do 
not trace well the direction of the minimum external tidal force, which implies the limitation of our definition of the void shape. 

It is interesting to see that our result is consistent with the $N$-body result of \citet{platen-etal08}, in spite of several technical 
differences. In their numerical work with a $N$-body simulation, the watershed void finder developed by \citet{platen-etal07} was 
used to identify the matter voids and the void inertia tensors were calculated in terms of the positions and masses of dark matter 
particles in the voids. Whereas in our observational analysis the HV02 algorithm was employed to find galaxy voids and the 
void inertia tensors were determined by using only the luminous void galaxies. The consistency of the observed signal of the 
void shape correlations presented here with their numerical result implies that the anisotropic spatial distributions of the void galaxies 
trace the overall non-spherical shapes of the matter voids. 

\section{COMPARISON OF THE OBSERVED SIGNAL WITH AN ANALYTIC MODEL}\label{sec:model}

We now compare the detected signal of the void shape correlations with an analytic model obtained as a 
modification of the halo shape correlations. Let us first briefly review the analytic models for the galaxy spin and shape 
correlations developed in the previous studies in the framework of the tidal torque theory.
\citet[][hereafter PLS00]{PLS00} developed the following analytic formula for the spatial correlations of the halo spin axes 
\begin{equation}
\label{eqn:eta_spin}
\eta_{\rm g,spin}(r) \equiv \langle\vert{\bf s}({\bf x})\cdot{\bf s}({\bf x}+{\bf r})\vert^{2}\rangle - \frac{1}{3} 
= \left[\frac{a_{\rm L}}{\sqrt{6}}\frac{\xi_{R_G}(r)}{\xi_{R_G}(0)}\right]^{2},\
\end{equation} 
where $\hat{\bf s}({\bf x})$ is the unit spin vector of a galactic halo, $\xi_{R_G}(r)$ represents the two-point correlation of 
the linear density field smoothed on the galactic scale of $R_{\rm G}\approx 0.5\,h^{-1}$Mpc, and $a_{\rm L}$ is the linear correlation 
coefficient that lies in the $[0,\ 3/5]$ according to the linear tidal torque theory 
\citep{dor70,white84,CT96,LP00,LP01,porciani-etal02a,porciani-etal02b}. PLS00 found fairly good agreement between Equation 
(\ref{eqn:eta_spin}) with an empirical value of $a_{\rm L}\approx 0.1$ and the observed galaxy spin correlations from the nearby universe 
\citep[see also][]{LE07,lee11}.

The critical aspect of Equation (\ref{eqn:eta_spin}), based on the linear tidal torque theory, is that the halo spin correlation drops 
very rapidly with $r$ as it is proportional to the square of the linear density correlation function. In other words, Equation 
(\ref{eqn:eta_spin}) predicts that the signal of the halo spin correlation is significant only on the scales of a few Mpc.
The follow-up numerical experiments found that the nonlinear evolution of the tidal shear field would induce larger scale 
correlations of the halo spin axes and suggested that the halo spin correlation should be described as a linear scaling of 
$\xi_{R_{\rm G}}$ \citep{catelan-etal01,LP08,HZ08}.  

The directions of the galaxy spin axes are hard to accurately determine in practice, expect for the case of the late-type spiral 
galaxies for which the thin disk approximation can be made.  Given that the angular momentum of a galactic halo is closely 
linked to the elongation of its ellipsoidal shape, the tidally induced spatial correlations are expected to be also present in 
the directions of the halo shapes that are readily observable in practice. Besides, unlike the halo spin axes that are more or 
less conserved after the decoupling of the halos from the tidal shear field,  the halo shapes do not retain their initial memory 
after the proto-halos decouple and thus their spatial correlations are likely to respond much more sensitively to the nonlinear 
growth of the tidal shear field than those of the halo spin axes.

\citet{lee-etal08} determined the halo shape correlations by analyzing the Millennium galaxy catalog and found them to be 
well approximated as a linear scaling of $\xi_{R_{\rm G}}$:
\begin{equation}
\label{eqn:eta_shape}
\eta_{\rm g,shape}(r) \equiv \langle\vert{\bf u}({\bf x})\cdot{\bf u}({\bf x}+{\bf r})\vert^{2}\rangle - \frac{1}{3} 
= {a_{\rm L}}\frac{\xi_{R_G}(r)}{\xi_{R_G}(0)}\, ,
\end{equation} 
where ${\bf u}$ represents the principal axes of the halo inertia tensors, which coincide with the directions of their ellipsoidal 
shapes. Note there are two differences between Equations (\ref{eqn:eta_spin}) and (\ref{eqn:eta_shape}). First, the halo shape 
correlations are proportional to the two-point linear density correlation function while the halo spin correlation is to the square of it, 
which indicate that the halo shapes possess larger-scale correlations than the halo spin axes. Second, the linear 
correlation coefficient $a_{\rm L}$ for the halo shape correlation is not reduced by a factor of $\sqrt{6}$, unlike the case of 
the spin correlation. This results from the fact that the halo minor axes are expected to be maximally correlated 
with the major principal axes of the tidal shear field \citep[e.g,][]{lee-etal08}, while the halo spin axes are only preferentially correlated 
with the intermediate tidal shear principal axes \citep{dor70,white84,CT96,LP00,porciani-etal02a,porciani-etal02b}. 
These two differences imply that the halo shapes should exhibit not only larger scale but also stronger correlations than 
the halo spin axes.   

Meanwhile, the validity of Equation (\ref{eqn:eta_spin}) was extended to the case of a void spin, a new concept, introduced by 
\citet{LP06} to quantify the effect of the tidal shear field on the dynamics of the void galaxies. Given the analytic result of 
\citet{lee06} that the linear correlation coefficient tends to have its maximum value in the regions with lowest local density, 
\citet{LP06} suggested that the spatial correlations of the void spin axes could be described by Equation (\ref{eqn:eta_spin}) 
with $a_{\rm L}=0.6$ (i.e., the maximum value of the linear correlation coefficient): 
\begin{equation}
\label{eqn:eta_vspin}
\eta_{\rm v,spin}(r) = \left[\frac{3}{5\sqrt{6}}\frac{\xi_{R_v}(r)}{\xi_{R_v}(0)}\right]^{2}\, ,
\end{equation} 
where $\xi_{R_v}$ is the two-point correlation function of the linear density field smoothed on the void scale that amounts 
to the effective Lagrangian radius of a void. Equation (\ref{eqn:eta_vspin}) was tested against a $N$-body simulation by \citet{LP06} 
and turned out to work well even though it has no fitting parameter.  
 
\begin{figure}
\includegraphics[scale=1.0]{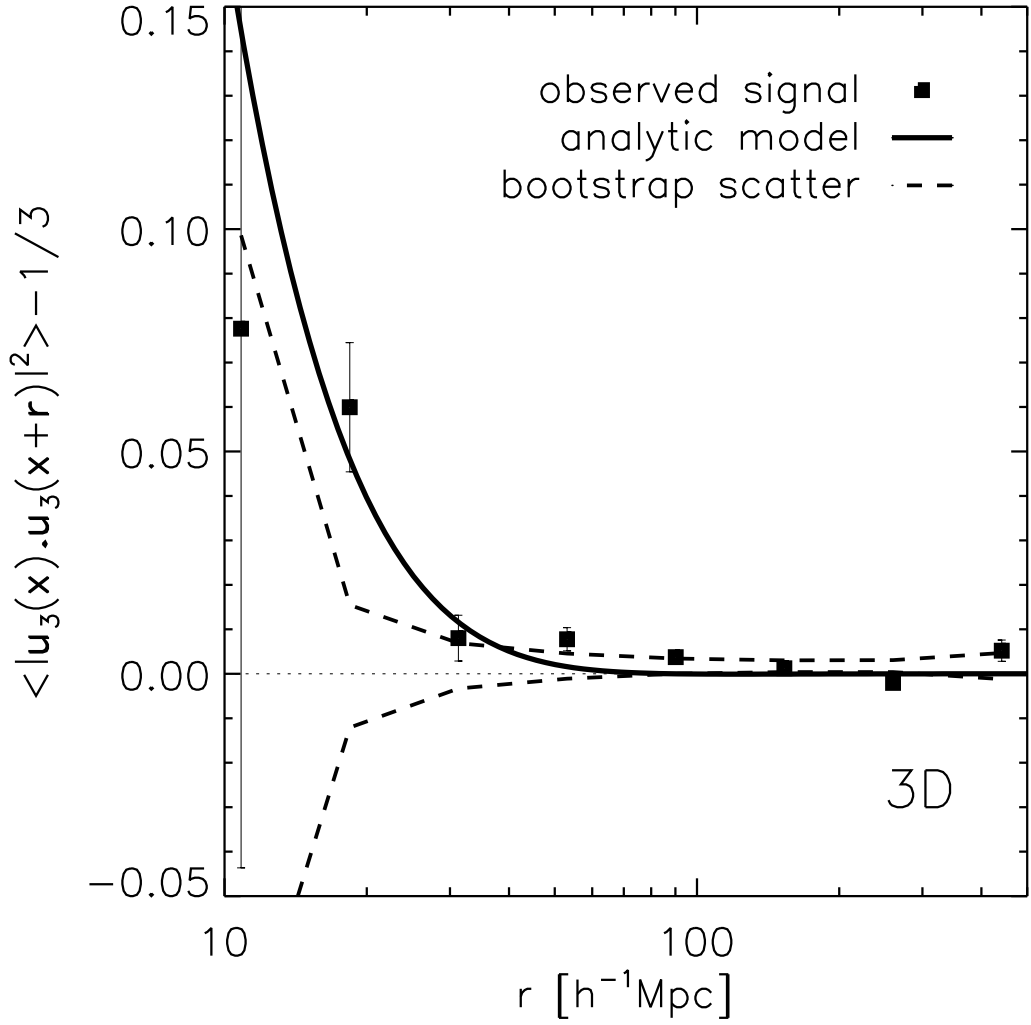}
\caption{Same as Figure \ref{fig:major} but for the case of the minor principal axes of the void inertia tensors. 
The solid line represents an analytic model based on the linear tidal torque theory.}
\label{fig:minor}
\end{figure}

\begin{figure}
\includegraphics[scale=1.0]{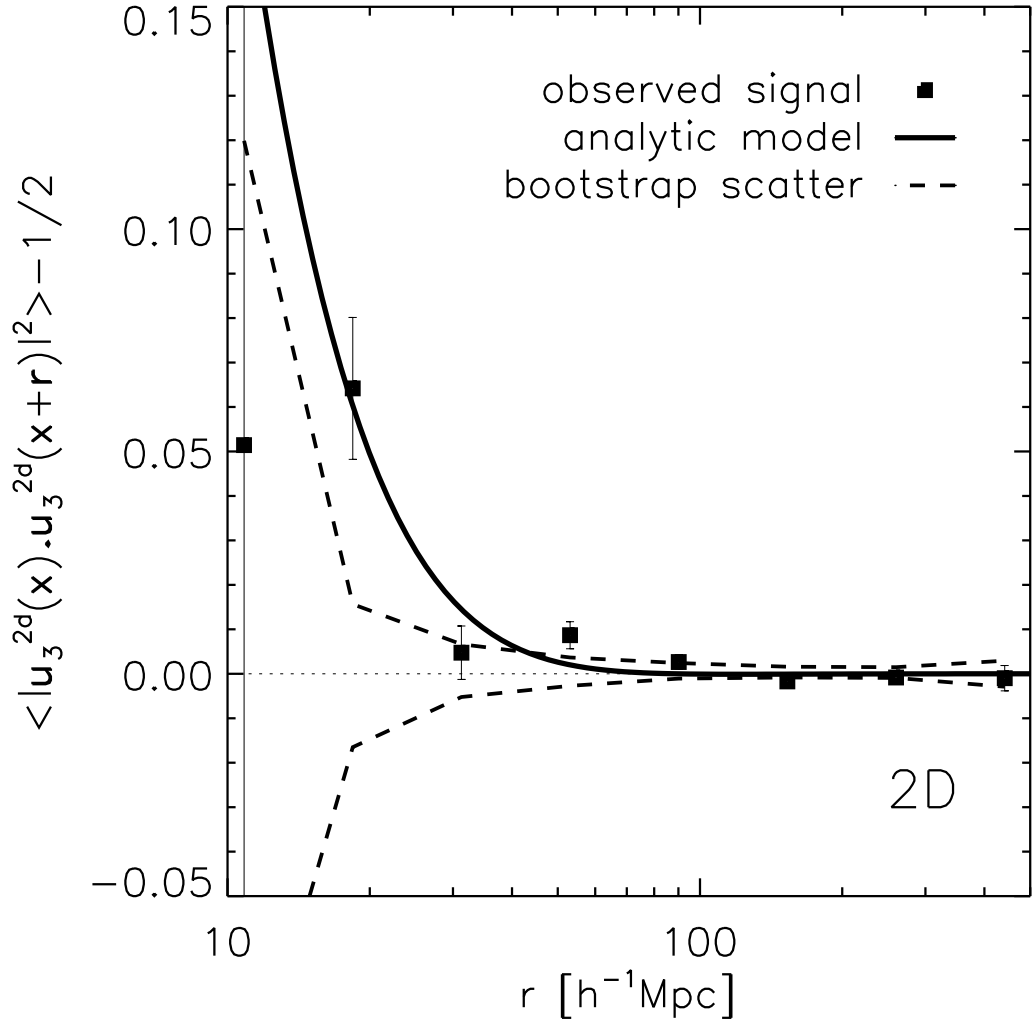}
\caption{Same as Figure \ref{fig:minor} but for the case of the projected minor principal axes onto the two dimensional 
plane of sky}
\label{fig:2d}
\end{figure}

In the light of the previous works reviewed here, we now suggest that the void shape correlation function 
$\eta_{\rm v}$ could be described by Equation (\ref{eqn:eta_shape}) with the maximum value of $a_{\rm L}=3/5$: 
\begin{equation}
\label{eqn:eta_vshape}
\eta_{\rm v,shape}(r) = \frac{3}{5}\frac{\xi_{R_v}(r)}{\xi_{R_v}(0)}\, . 
\end{equation}
For the case of the spatial correlation of the projected void shapes, the proportionality constant changes from $3/5$ to $4/5$ 
\citep[see Appendix H in][]{LP01}
\begin{equation}
\label{eqn:eta_vshape2}
\eta_{\rm v}(r) = \frac{3}{4}\frac{\xi_{R_v}(r)}{\xi_{R_v}(0)}\, . 
\end{equation}
The solid lines in Figures \ref{fig:minor} and \ref{fig:2d} represent Equations (\ref{eqn:eta_vshape}) and (\ref{eqn:eta_vshape2}), 
respectively, compared with the numerical results (dots) obtained in section \ref{sec:signal}. 
As can be seen, the strengths of the correlation signals detected at the distance bin of $r\approx 20\,h^{-1}$Mpc shown in 
Figures \ref{fig:minor}-\ref{fig:2d} match well the analytic predictions of Equations (\ref{eqn:eta_vshape})-(\ref{eqn:eta_vshape2}). 

\section{CROSS-CORRELATIONS BETWEEN THE VOID SHAPES AND THE VELOCITY SHEARS}
\label{sec:shear}

The observational signal of the void shape correlation and its good agreement with the analytic model based on 
the tidal torque theory provides compelling {\it indirect} evidence for the dominant effect of the tidal shear field on void 
shapes. It would be tantalizing to detect a {\it direct} observational evidence for the connection between the void 
shapes and the tidal shears.  If the tidal shear forces drive the voids to develop non-spherical shapes, then it is expected 
that the void shapes are least stretched in the directions of the maximum gravitational force.  
In other words, the minor axes of the void inertia tensors are expected to not be isotropically oriented 
in the principal frame of the local tidal tensors but preferentially aligned with the major principal axes of the local tidal tensors. 

The tidal shear is identical to the velocity shear if the peculiar velocity is calculated as a gradient of the gravitational potential, under the 
assumption of no vorticity. 
\citet{lee-etal14} linearly reconstructed the velocity shear field in the nearby universe over a volume of $180^{3}\,h^{-3}$Mpc$^{3}$ that 
corresponds to the redshift range of $0\le z\le 0.08$. 
The local volume was divided into the $256^{3}$ cubic grids, in each the velocity shear tensors $\Sigma_{ij}({\bf x})$ 
were calculated by analyzing the peculiar velocity field linearly reconstructed by \citet{wang-etal12} from the samples of the galaxies 
with masses larger than $10^{12}\,h^{-1}M_{\odot}$ from the SDSS DR7.
 
By conducting the Fourier-transformation of the peculiar velocity field ${\bf v}({\bf x})$ into the Fourier space, \citet{lee-etal14} 
calculated the Fourier amplitude, $\tilde{\Sigma}_{ij}({\bf k})$, of the velocity shear field smoothed by a Gaussian window on the 
filtering scale of $R_{f}$  according to the definition of \citet{hoffman-etal12} 
\begin{equation}
\tilde{\Sigma}_{ij}({\bf k}) = -\frac{1}{2H_{0}}
\left(k_{i}\tilde{v}_{j}+k_{j}\tilde{v}_{i}\right)\exp\left(-\frac{R^{2}_{f}k^{2}}{2}\right)\, , 
\end{equation} 
where $\tilde{{\bf v}}({\bf k})=(\tilde{v}_{1},\ \tilde{v}_{2},\ \tilde{v}_{3})$ is the Fourier amplitude of the peculiar velocity 
field.  Then, they conducted the inverse Fourier transformation of $\tilde{\Sigma}_{ij}({\bf k})$ to determine the velocity shear 
tensor $\Sigma_{ij}({\bf x})$ at each grid point, whose three eigenvectors, corresponding to three eigenvalues, were obtained via 
the similarity transformation.  For the detailed description of the linear reconstruction of the velocity shear field, see 
\citet{lee-etal14}. 
  
Among the $843$ voids in the sample, we find that only $220$ voids are located within the local volume of 
$180^{3}\,h^{-3}$Mpc$^{3}$ where the velocity shear field was reconstructed. 
The grid in which the luminosity center of each selected void belongs is identified and we define the major, intermediate and minor 
principal axes of the velocity shear tensor of the grid as its eigenvectors corresponding to the largest, second largest and 
smallest eigenvalues. The smoothing scale of the velocity shear tensor is set at $R_{f}=5\,h^{-1}$Mpc that is close to the mean 
value of the effective Lagrangian radius of the selected voids \citep{pan-etal12}. 

Before measuring the alignments between the minor axes of the void inertia tensors and the principal axes of the 
local velocity shear tensors, however, we would like to examine whether or not the principal axes of both of the tensors 
are isotropic in their orientations with respect to the LOS directions. As is well known, the observational datasets 
measured in the redshift space are likely to be contaminated by LOS systematics. A false signal of alignment between the principal 
axes of the two tensors could be produced if the LOS systematics contaminated the measurements of the two tensors and in 
consequence induced preferential alignments (or anti-alignments) between the (LOS) directions and their principal axes. 

\begin{figure}[tb]
\includegraphics[scale=1.0]{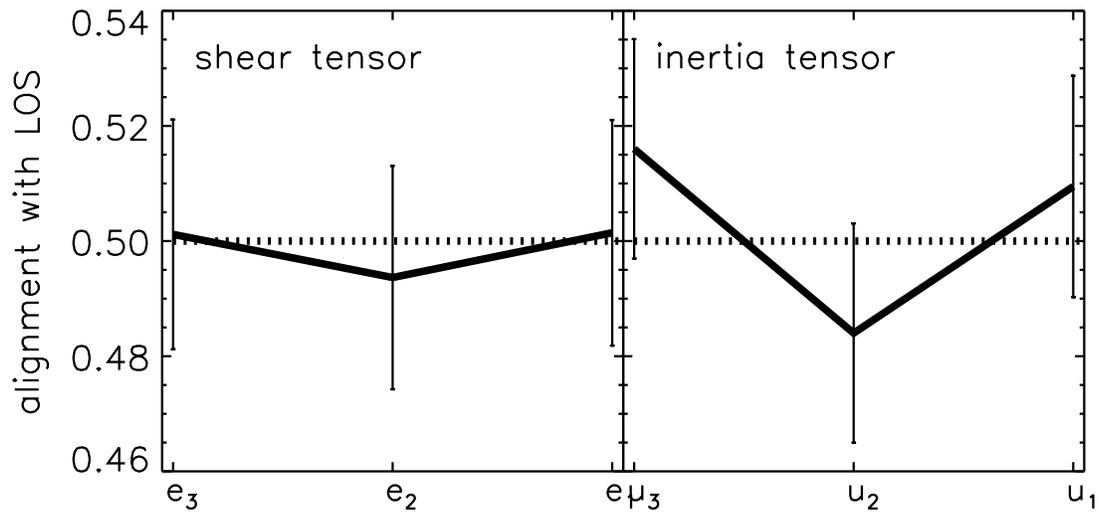}
\caption{Mean cosines of the angles between the principal axes of the velocity shear (void inertia) tensors and the 
line of sight directions at the centers of the samples voids in the left (right) panel.}
\label{fig:los}
\end{figure}

We measure the cosine of the angle between the three principal axes of the velocity shear tensors ${\bf e}_{i}$ and LOS direction, 
$\hat{\bf h}_{\rm los}$, at the luminosity center of  each void  and then take the average value over the $230$ voids located in the local 
volume as  $\langle\vert{\bf e}_{i}\cdot{\bf h}_{\rm los}\vert\rangle$. If the shear principal axes are isotropic in their orientatons relative to 
the LOS directions, then, the average value of the cosines of the angles would be $0.5$ for all of the three cases of 
${\bf e}_{1},\ {\bf e}_{2},\ {\bf e}_{3}$. The left panel of Figure \ref{fig:los} plots the results, the errors represent one standard 
deviation in the measurement of the average.  As can be seen, there is no statistically significant signal of alignment between ${\bf e}_{i}$ 
and ${\bf h}_{\rm los}$.  The differences between the resulting average values and the expectation value of $0.5$ are all within the 
statistical errors, indicating that the principal axes of the velocity shear tensors at the luminosity centers of the selected voids are 
isotropic in their orientations with respect to ${\bf h}_{\rm los}$.  We repeat the same calculation but using the principal axes of the void 
inertia momentum tensors, ${\bf u}_{i}$, and find that their orientations are also isotropic relative to ${\bf h}_{\rm los}$, as shown in the 
right panel of Figure \ref{fig:los}. 

Having confirmed that there is no significant LOS systematics in the measurements of the principal directions of the void inertia and 
local tidal shear tensors, we now measure the cosine of the angle between the minor principal axis of the void inertia tensor and the 
three principal axes of the velocity shear tensor at the luminosity center of each selected void as 
$\cos\theta = \vert{\bf u}_{3}\cdot{\bf e}_{i}\vert$ for $i=1,\ 2,\ 3$. 
Binning $\cos\theta$ and counting the voids whose values of $\cos\theta$ belong to a given bin, we evaluate the 
probability density distribution, $p(\cos\theta)$, for $i=1,\ 2,\ 3$. 
If the minor principal axes of the void inertia tensors are aligned (anti-aligned) with the principal axis of the shear tensor, 
then $p(\cos\theta)$ will show an increase (decrease) with $\cos\theta$. 

\begin{figure}
\includegraphics[scale=1.0]{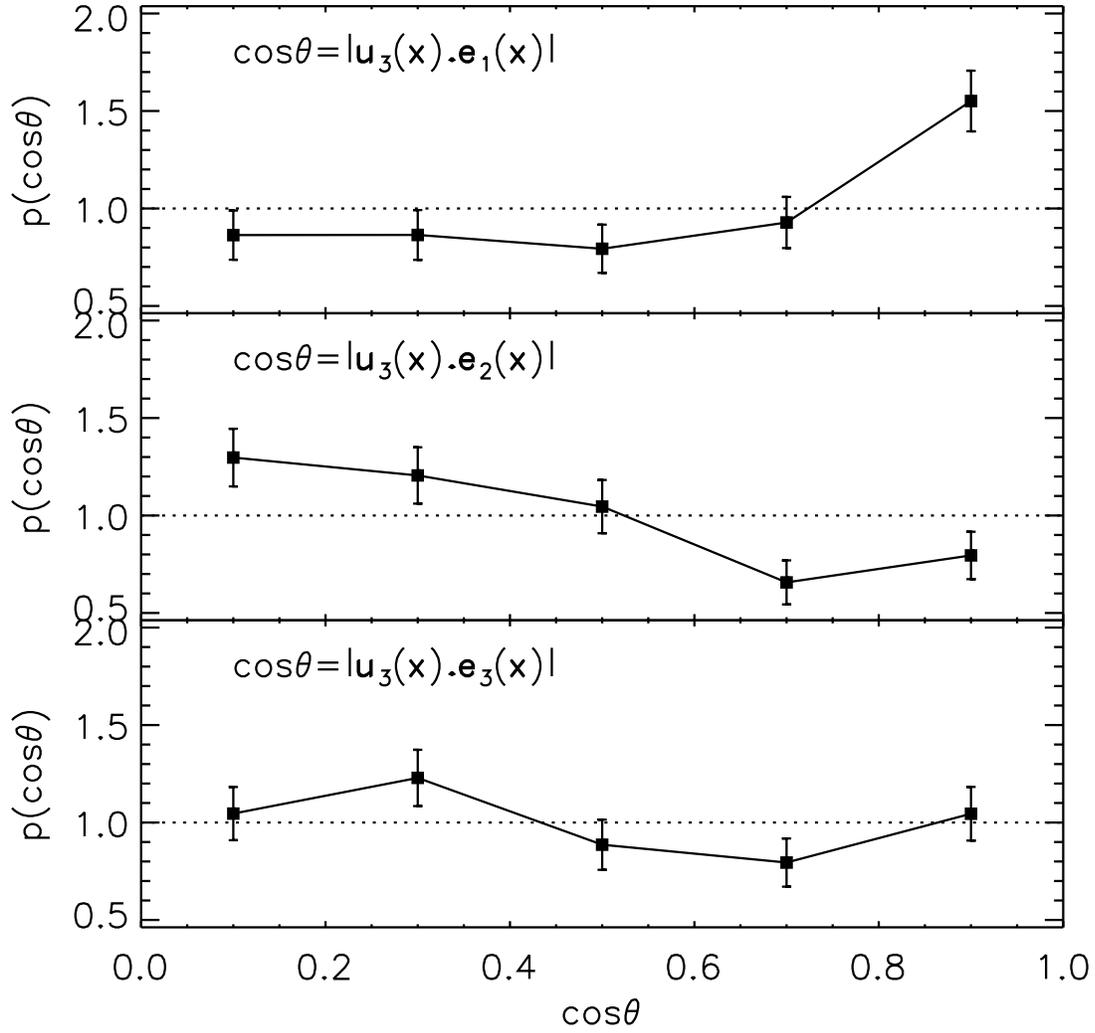}
\caption{Probability density distributions of the cosines of the angles of the minor principal axes of the void inertia tensors 
relative to the major, intermediate, and minor principal axes of the velocity shear tensors in the top, middle, and 
bottom panels, respectively. }
\label{fig:pro3}
\end{figure}

Figure \ref{fig:pro3} shows the results as square dots with Poisson errors. As can be seen, the minor principal axes of the void inertia 
tensors ${\bf u}_{3}$ exhibit a tendency of alignment (anti-alignment) with the major (intermediate) principal axes of the velocity shear 
tensor ${\bf e}_{1}$ (${\bf e}_{2}$), providing a {\it direct} observational evidence for the theoretical claim that the void shapes are 
nonspherical and least stretched along the direction of the maximum tidal force. Note also that no signal of alignment nor 
anti-alignment is found between ${\bf u}_{3}$ and ${\bf e}_{3}$. 

\begin{figure}
\includegraphics[scale=1.0]{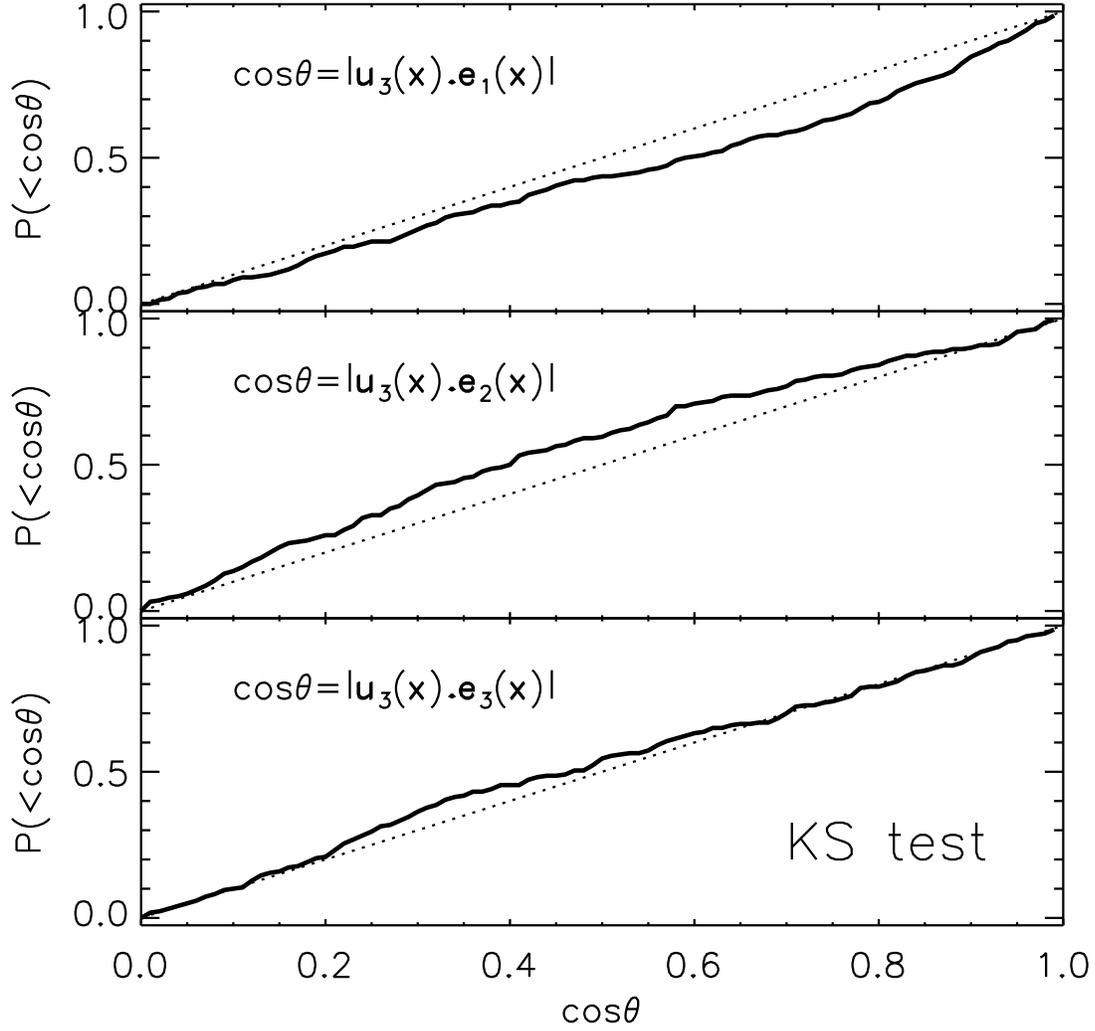}
\caption{KS tests of the null hypothesis (dotted line) of no correlation between the void inertia and the velocity shear tensors. 
The larger the maximum difference between the cumulative probability, $P(<\cos\theta)$, from the observational data (solid line) 
and the expectation of the null hypothesis, $P(<\cos\theta)=\cos\theta$, is,  the larger confidence the null hypothesis is rejected at.}
\label{fig:kstest}
\end{figure}

Given that the number of the voids, $220$, is rather small, we perform the KS test of the null hypothesis of no alignment. If there were 
no alignments between the principal axes of the two tensors, then the probability $p(\cos\theta)$ would be unity, or equivalently 
the cumulative probability, $P(<\cos\theta)$, would equal $\cos\theta$.  Comparing this expectation value of $\cos\theta$ with 
the observed cumulative probabilities $P(<\cos\theta)$, we find the largest differences between the two, 
$\vert P(<\cos\theta)-\cos\theta\vert$ to be $0.120,\ 0.121,\ 0.075$ for the cases of ${\bf e}_{1},\ {\bf e}_{2},\ {\bf e}_{3}$, respectively.  
The results are shown in Figure \ref{fig:kstest}.  According to the KS one sample test with $220$ data points \citep{WJ12},  
the null hypotheses of no alignment between ${\bf u}_{3}$ and ${\bf e}_{1}$ and no anti-alignment between ${\bf u}_{3}$ and 
${\bf e}_{2}$ are rejected at the $99.8\%$ and $99.9\%$ confidence levels, respectively.  

\begin{figure}
\includegraphics[scale=1.0]{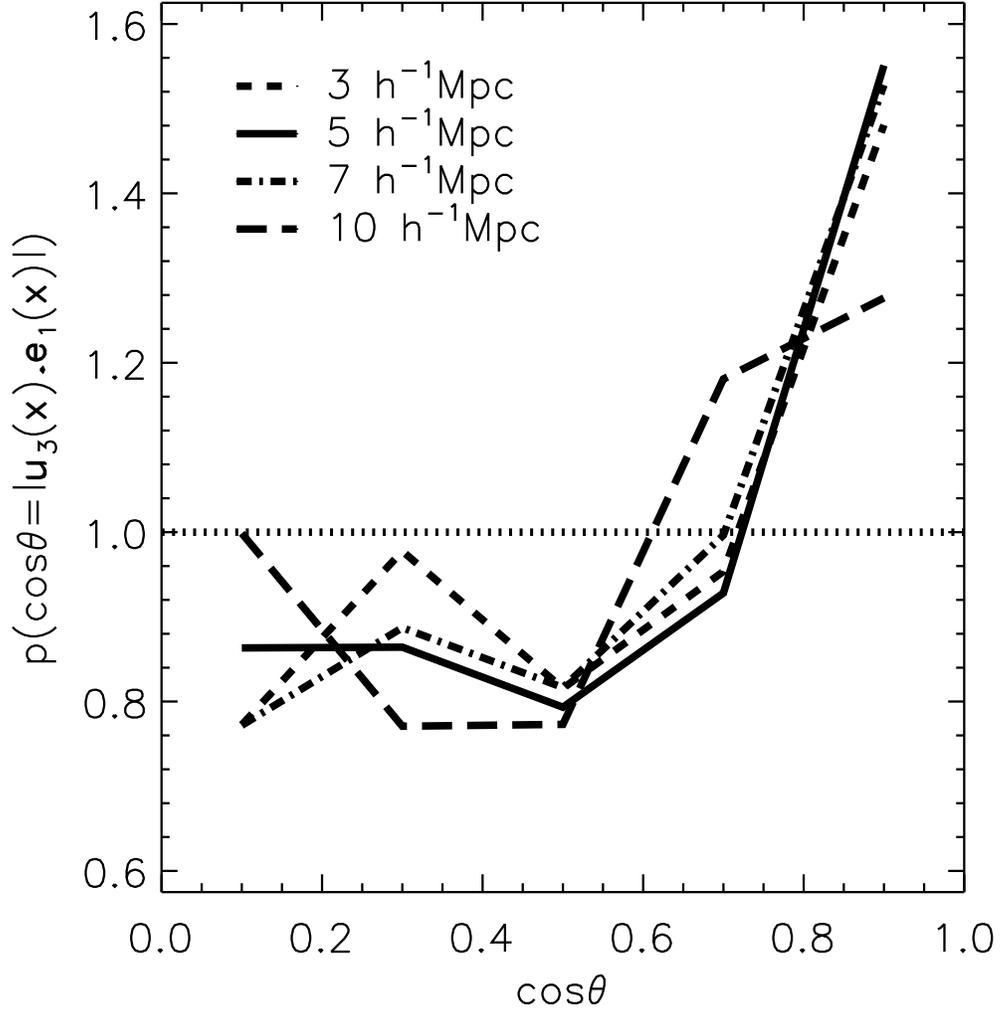}
\caption{Probability density distributions of the cosines of the angles between the minor principal axes of the void inertia tensors 
and the major principal axes of the velocity shear tensors for four different cases of the filtering scales.}
\label{fig:filter}
\end{figure}

We repeat the whole process described above for three different filtering scales: $R_{f}=3,\ 7,\ 10\,h^{-1}$Mpc. Figure \ref{fig:filter} 
shows how the probability density distributions of the cosines of the angles between ${\bf u}_{3}$ and ${\bf e}_{1}$ change with $R_{f}$. 
As can be seen, the distribution of $p(\cos\theta)$ does not show a large difference in its behaviour among the four different cases of 
$R_{f}$. But, we also note that the increment of $p(\cos\theta)$ with $\cos\theta$ becomes shallower when $R_{f}=10\,h^{-1}$Mpc. Although 
it would be interesting to investigate at what scale of $R_{f}$ the alignment tendency would disappear, this issue cannot be addressed 
in the current work because the validity of the reconstructed velocity shear field will break down when the smoothing scale becomes 
larger than $10\,h^{-1}$Mpc due to the boundary effect \citep{wang-etal12}.

 \section{SUMMARY AND DISCUSSION}
\label{sec:summary}
 
We have measured the spatial correlations of the principal axes of the inertia tensors of the galaxy voids from the SDSS DR7 
and detected a clear signal as high as $4\sigma$ for the case of the minor principal axes on the scale of $20\,h^{-1}$Mpc.  The 
observed signal has been found to be robust against the projection of the principal axes of the void inertia tensors onto the plane 
of sky, verifying that the redshift-space distortions have little contamination effect on the directions of the minor principal axes of 
the void inertia tensors, which is consistent with the previous works \citep[e.g., see][]{FN09}. By constructing a simple analytic model 
with no fitting parameter for the void shape correlation function, based on tidal torque theory, and by comparing the observed signal with 
the analytic model, we have shown that the observed signal matches the predicted value fairly well.  

We have also presented a direct observational evidence for the anisotropic orientations of the void shapes in the principal 
frame of the local velocity shear tensors that were linearly reconstructed from the peculiar velocity field obtained from the galaxy 
group catalog of the SDSS DR7. It has been found that the minor principal axes of the void inertia tensors are indeed preferentially 
aligned with the major principal axes of the local velocity tensors. Since the tidal shear field is identical to the linearly reconstructed 
velocity shear field, this result agrees well with the theoretical expectation that the void shapes stretch least in the directions of the 
maximum tidal torque forces. The alignment tendency is robust against the change of the filtering scale on which the 
local velocity field is smoothed, albeit showing a mild decrease as $R_{f}$ reaches up to $10\,h^{-1}$Mpc. 
Performing the KS test, we reject the null hypothesis of no correlation between the void shapes and the local tidal shear tensors
at $99.9\%$ confidence level.

Given the rescaled linear density correlation function $\xi(r)/\sigma$ is almost independent of the normalization amplitude 
$\sigma_{8}$ but still sensitive to the matter density parameter $\Omega_{m}$, we suggest that the void shape correlation 
function should be useful to break the $\Omega_{m}-\sigma_{8}$ degeneracy when combined with the other LSS statistics.  
The void shape correlation function is also expected to provide an independent constraint on the neutrino mass $\mu_{\nu}$ 
since the behavior of $\xi(r)/\sigma$ would change with an increase of $m_{\nu}$ in the large-scale tail.  
However, the detection of the void shape correlation has only been successfully detected at one distance bin of 
$20\,h^{-1}$Mpc due to the limited numbers of the low-$z$ voids, and thus should be regarded as marginal. 
To use the void shape correlation as a probe of cosmology,  however, it will be necessary to overcome the small number statistics 
by exploring the high-$z$ voids and tracing the evolution of the void shape correlations.  Our future work is in this direction.

\acknowledgments

We thank an anonymous referee for helpful comments.
This work was initiated during the workshop on "Cosmic Voids in the Next Generation of Galaxy Surveys" held at the Ohio State
University's Center for Cosmology and Astro-Particle Physics from August 18-20, 2014. We thank P. Sutter and the other organizers 
of the workshop for making interactactions between participants possible during the workshop. 
J.L. thanks H. Wang for providing the dataset of the reconstructed peculiar velocity field. J.L. was supported by the 
research grant from the National Research Foundation of Korea to the Center for Galaxy Evolution Research  
(NO. 2010-0027910) and the grant by the Basic Science Research Program through the National Research 
Foundation of Korea (NRF) funded by the Ministry of Education (NO. 2013004372). 
FH acknowledges the support of a PUCE research grant.

\clearpage
\clearpage
\clearpage
\clearpage
\clearpage
\clearpage
\clearpage
\clearpage
\end{document}